\documentclass[11pt,twoside]{article}  
\usepackage{apn3conf}

\def\noi{\noindent}
\def\be{\begin{equation}}
\def\ee{\end{equation}}

\begin{document}   

%
%
%
%

\title{Stellar Rotation and the Formation of Asymmetric Nebulae}

%
%
%

\author{Vikram V. Dwarkadas } 
\affil{ASCI FLASH Center, U Chicago, 5640 S.~Ellis Ave, Chicago IL 60637 }

%
%

\contact{Vikram Dwarkadas }
\email{vikram@flash.uchicago.edu }

%
%
%
%
%

\paindex{Dwarkadas, V. V. }

%
%

\authormark{Dwarkadas, V.~V. }

%
%



\begin{abstract}          
We illustrate how rotation of the central star can give rise to
latitudinal variations in the wind properties from the
star. Interaction of these winds with the surrounding medium can
produce asymmetrical planetary nebulae.
\end{abstract}

%
%
\vspace{-0.25truein}
\section{Introduction }

Planetary Nebulae (PNe) exhibit a dazzling variety of shapes, as
emphasized so effectively in various talks at this conference. While
this diversity does not allow easy classification, it is universally
accepted that most, if not all, PNe do not show spherical
symmetry. Balick (1987) classified them as ranging from spherical
through elliptical to bipolar, which show two lobes emanating from an
equatorial waist. The origin of this asymmetry in shape has been the
cause of much speculation. A favored model for many years was the
Generalized Interacting Stellar Winds Model (GISW; see Frank 1999 and
references therein), which stated that the asymmetry was due to the
expansion of the nebula within a structured ambient medium, whose
density was higher at the equator than at the poles. The high
equatorial density inhibits the expansion at the equator, leading to a
prolate or, for very high density contrasts, a bipolar nebula.

While many authors have shown (eg.~Dwarkadas, Chevalier \& Blondin
1996) that this model can reproduce a vast diversity of morphologies,
it does not seem to reproduce many of the details. Besides, the
nagging question of what produces the asymmetry in the surrounding
medium has always persisted. Rotation, binary evolution, magnetic
fields, pre-existing disks, stellar pulsations and many other
suggestions have been put forward, none of them ubiquitous or totally
convincing. Many authors have also questioned whether it is likely
that such an asymmetric density distribution is really present in
every aspherical planetary nebula, given the lack of visible evidence.

In this work we show that rotation of the central star can lead to
aspherical mass-loss from the star. This aspherical wind expanding
into a constant density medium forms an aspherical nebula, without
having to resort to any external asymmetries. Our results, although
derived primarily from radiatively driven wind-theory applied to high
luminosity stars (see Dwarkadas \& Owocki 2002 for further details),
have more general applicability, and could prove relevant for the
central stars of PNe.

\section{Effect of Rotation}

For a rotating star, reduction by the radial component of the
centrifugal acceleration yields an effective gravity that scales with
co-latitude $\theta$ as

\be
g_{eff}(\theta) = g \, \left [ 1 - \kappa_e F/gc - \Omega^2 \sin^2 \theta
\right ] \, ,
\ee
\noi
where $\Omega \equiv \omega/\omega_c$, with $\omega$ the star's
angular rotation frequency, and $\omega_c \equiv \sqrt{g/R}$, $F$ is the 
stellar radiative flux and $\kappa_e$ is the electron scattering opacity. 

For a broad range of stellar wind-driving mechanisms, flow terminal
speed scales directly with the surface escape speed. If the star is
uniformly bright, then the latitudinal variation of terminal speed
goes as:
\be
{v_\infty (\theta) \over v_\infty (0) } =
\left [ {g_{eff} (\theta) \over g_{eff} (0) } \right ]^{1/2} =
\left [ 1 - {\Omega^2 {{\sin}^2\theta}
\over 1 - \Gamma } \right ]^{1/2} \, ,
\ee
where the polar speed 
$v_\infty (0) \sim \sqrt{gR(1-\Gamma)}$, and $\Gamma =$ Eddington 
parameter.

However, if $F(\theta) \propto g (1 - \Omega^2 \sin^2 \theta)$
(gravity darkening effect, von Zeipel, 1924) then this implies
\be
{v_\infty (\theta) \over v_\infty (0) } =
\left [ {g_{eff} (\theta) \over g_{eff} (0) } \right ]^{1/2} =
\left [ 1 - {\Omega^2 {{\sin}^2\theta}
} \right ]^{1/2} \, .
\ee

The evolution of the wind velocity is quite general, and does not
depend strongly on the details of the wind-driving mechanism. This is
not true for the mass-loss rate. In order to determine the mass-loss
rate, we must include more details of the specific wind-driving
mechanism. We derive our results from radiatively-driven wind theory
(see Owocki et al.~1998; Dwarkadas \& Owocki 2002), applicable mainly
to PNe with hot O and WR type central stars.

For a star with luminosity $L$, the CAK, line-driven mass loss rate
(Castor, Abbott \& Klein 1975) can be written in terms of the mass
flux ${\dot m} \equiv {\dot M}/4 \pi R^2$ at the stellar surface
radius $R$, which then depends on the surface radiative flux $F=L/4
\pi R^2$ and the effective surface gravity $g_{eff} \equiv
(GM/R^2)(1-\Gamma)$ through (Owocki, Cranmer \& Gayley 1998; Dwarkadas
\& Owocki 2002)
\be
    {\dot m} \propto F^{1/\alpha} ~ g_{eff}^{1-1/\alpha} \, .
\ee
\noi
where $\alpha < 1$. If the radiative flux $F$ is constant over the
stellar surface, then $\kappa_e F/gc= \Gamma$ in equation (1), and
application of equation (4) yields

\be
{{\dot{m}(\theta)} \over {\dot{m}(0)}} =
\left [ {g_{eff} (\theta) \over g_{eff} (0) } \right ]^{1 - 1/\alpha} =
\left [ 1 - {\Omega^2 {{\sin}^2\theta}
\over 1 - \Gamma } \right ]^{1 - 1/\alpha} \, .
\ee
\noi
Since the exponent $1-1/\alpha$ is negative, the mass flux from such a
uniformly bright, rotating star increases from the pole ($\theta = 0$)
to the equator ($\theta = 90$). 

However, taking gravity darkening into account yields the mass flux
scaling:

\be
\label{eq:mfvz}
{{\dot{m}(\theta)}  \over {\dot{m}(0)}} =
{F (\theta) \over F(0) } = 1 - {\Omega}^2 {{\sin}^2\theta}
\ee
\noi
i.e.~the mass flux is highest at the poles, and decreases towards the equator.

Rotation thus generates a wind that is faster at the poles, but denser
at the equator. The inclusion of gravity darkening leads to a wind
that is both faster and denser at the poles than the equator. These
effects are illustrated in Fig 1. 

\begin{figure}
\epsscale{0.85}
\plotone{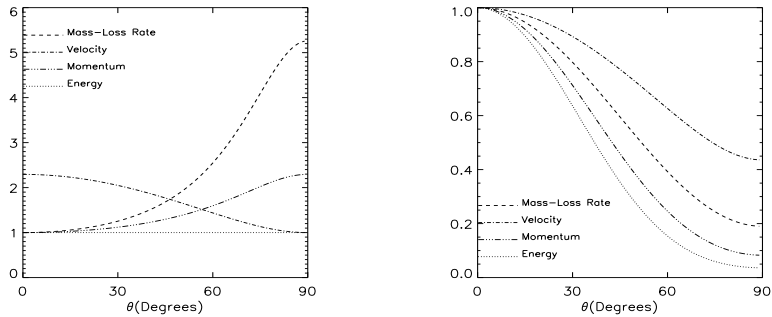}
\caption{The variation of the wind velocity, mass-loss rate, momentum
and energy with co-latitude, for the case of rotation with (right) and
without (left) gravity darkening, using $\alpha$=0.5. $\theta=0$
refers to pole. }
\end{figure}

Figure 2 shows results from simulations for values of the rotation
parameter ${\tilde{\Omega}}=\Omega / (1 - \Gamma)$, without including
gravity darkening. The nebula is still in the early stages of
evolution, and there is no hot, high-pressure bubble driving the
expansion. As the rotation parameter increases from 50\% to 90\% of
critical, the nebular morphology changes from nearly spherical to
distinctly bipolar.

Inclusion of gravity darkening alters only the interior density
distribution, and not the overall morphology. This indicates that it
is the wind velocity distribution that is primarily responsible for
shaping the nebula.

\begin{figure}
\plotone{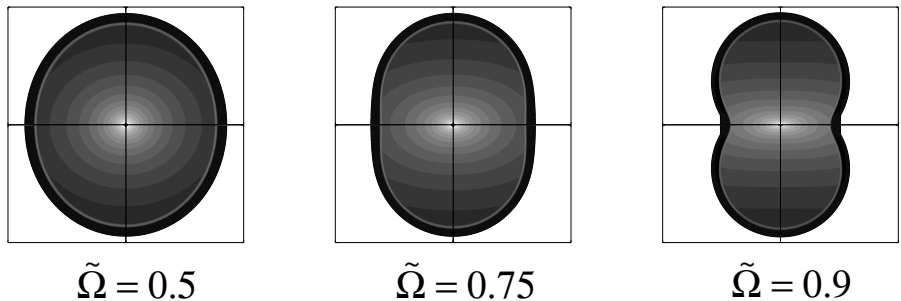}
\caption{ Density contours from simulations of an aspherical wind from
a rotating star expanding in a constant-density medium.  }
\end{figure}

\section{Conclusions and Discussion}

We have shown that rotation can significantly modulate the winds from
stars, leading to higher velocities and larger wind-momentum at the
poles as compared to the equator. This can drive an aspherical, and
even bipolar wind-blown nebula, without having to resort to any
external, ad-hoc density asymmetry. The nebula will start out as a
momentum-driven, aspherical, structure, whose morphology depends on
the rate of rotation. Over time it will slowly {\it lose} its
asphericity, becoming more and more spherical as it evolves into an
energy-conserving bubble. This situation is inverse to that of the
GISW model, where the nebula slowly becomes more aspherical over time.

The results derived herein for the mass-flux are obtained using
radiatively-driven wind theory from hot, luminous stars. However we
emphasize that the shaping of the nebula depends mainly on the
asymmetry in velocity, which is not strongly dependent on the specific
wind-driving mechanism. A star rotating at a large fraction of
critical velocity will be flattened, resulting in a faster wind at the
poles than the equator, producing an elliptical or bipolar nebula.

The question remains then whether such large rotation velocities are
typical of PNe central stars. AGB stars in general are known to be
slow rotators. However we also know of exceptions such as V Hydra
(Barnbaum et al.~1995), which is claimed to be rotating at close to
critical velocity. Dorfi and Hoefner (1996) have shown that even small
rotational velocities (of the order of 2 km/s) at the stellar
photosphere can cause significant variations in the outflow velocities
and mass-loss rate. The variations that they find in their models look
similar to those in our models, although they are computed for
dust-driven winds.

The presence of a binary companion can lead to increased rotation
rates. Common envelope binary evolution, the presence of planets
around stars and tidal spin-up by a binary companion all tend to
increase the rotation velocities of stars.  Many authors (see Bond
2000) have found that a large percentage of PNe central stars may
exist in binary systems. In view of the theoretical possibility, as
well as presently available observational evidence (however slim), we
feel that fast rotation of PNe central stars cannot be ruled out.

\acknowledgements Vikram Dwarkadas is supported by Award \#
AST-0319261 from the National Science Foundation, and by the
US.~Dept.~of Energy grant \# B341495 to the ASCI Flash Center (U
Chicago). Collaboration with Dr.~Stan Owocki on this research is
gratefully acknowledged. A big thanks to the organisers for inviting
me to an engrossing conference in a fascinating setting.

\vspace{-0.05truein}
%
%
%
%



\begin{references}
\reference Balick, B.~1987, AJ, 94, 671
\reference Barnbaum, C., Morris, M., \& Kahane, C.~1995, ApJ, 450, 862
\reference Bond, H.~E.~2000, in APN2, ASP Conf Series 199, eds.J. H. Kastner, 
N. Soker, and S. Rappaport., (San Francisco: ASP), 115 
\reference Castor, J. I., Abbott, D. C., \& Klein, R. I.~1975, ApJ, 195, 157
\reference Dorfi, E. A., \& Hoefner, S~1996, A\&A, 313, 605
\reference Dwarkadas, V.~V., Chevalier, R.~A., \& Blondin J.~M.~1996, ApJ, 
457, 773
\reference Dwarkadas, V.~V., \& Owocki, S.~2002, ApJ, 581, 1337
\reference Frank, A.~1999, NewAR, 43, 31
\reference Owocki, S.~P., Cranmer, S.~R., \& Gayley, K.~G.~1998, ApSS, 260, 
1490
\reference von Zeipel, H.~1924, MNRAS, 84, 684




\end{references}
\end{document}